\documentclass[twocolumn,trackchanges]{aastex631}

\usepackage{amsmath}
\usepackage{booktabs}
\usepackage{color}
\hypersetup{
 colorlinks=true,%
 linkcolor=black,
 citecolor=blue,
}
\def\MS#1#2{\textcolor{black}{#2}}

\def\HK#1#2{\textcolor{black}{#2}}
\def\HKK#1#2{\textcolor{black}{#2}}
\def\HKKK#1#2{\textcolor{teal}{#2}}

\def\RJM#1#2{\textcolor{black}{#2}}

\begin{document}

\title{Can the solar p-modes contribute to the high-frequency transverse oscillations of spicules?}

\author[0000-0003-1134-2770]{Hidetaka Kuniyoshi}
\affiliation{Department of Earth and Planetary Science, The University of Tokyo,
7-3-1 Hongo, Bunkyo-ku, Tokyo, 113-0033, Japan}

\author[0000-0002-7136-8190]{Munehito Shoda}
\affiliation{Department of Earth and Planetary Science, The University of Tokyo,
7-3-1 Hongo, Bunkyo-ku, Tokyo, 113-0033, Japan}

\author[0000-0001-5678-9002]{Richard J. Morton}
\affiliation{Northumbria University, Newcastle upon Tyne, NE1 8ST, UK}

\author[0000-0001-5457-4999]{Takaaki Yokoyama}
\affiliation{Astronomical Observatory, Kyoto University, Sakyo-ku, Kyoto, 606-8502, Japan}



\begin{abstract}

\HK{}{Lateral motions of} spicules serve as vital indicators of transverse waves in the solar atmosphere, and their study is crucial for understanding the wave heating process of the corona. Recent observations have focused on "high-frequency" transverse waves (periods $<100 \ \rm s$), which have the potential to transport sufficient energy for coronal heating. These high-frequency spicule oscillations are distinct from granular motions, \RJM{having}{which have} much longer \RJM{periods}{time scales} of $5$--$10 \ \rm min$. Instead, it is proposed that they are generated through the mode conversion from high-frequency longitudinal waves that arise from a shock steepening process. \HK{While longitudinal waves can be generated by both granular motions and p-modes, our study utilized two-dimensional magneto-convection simulations to investigate the contribution of p-modes to the generation of high-frequency spicule oscillations.}{Therefore, these oscillations may not solely be produced by the horizontal buffeting motions of granulation but also by the leakage of p-mode oscillations. To investigate the contribution of p-modes, our study employs a two-dimensional magneto-convection simulation spanning from the upper convection zone to the corona. During the course of the simulation, we introduce a p-mode-like driver at the bottom boundary.} We reveal a notable increase in the mean velocity amplitude of the transverse oscillations in spicules, ranging from $10 \%$ to $30 \%$, and attribute this to the energy transfer from longitudinal to transverse waves. 
This effect results in an enhancement of the estimated energy flux by $30$--$80 \%$.

\end{abstract}



\section{Introduction} \label{sec:intro}

The temperature of the solar corona surpasses $10^6$ K, exceeding the surface temperature by several hundredfold \citep{Grotrian_1939, Edlen_1943_ZAP}. 
The underlying cause for the high-temperature corona is believed to reside within the solar magnetic field. However, the precise mechanisms responsible for this magnetically-driven heating continue to be a topic of active scientific debate, referred to as the ``coronal heating problem" \citep{Klimchuk_2006_SoPh, Klimchuk_2015_RSPTA, Reale_2014_LRSP, DeMoortel_2015_RSPTA, Cranmer_2019_ARAA, VanDoorsselaere_2020_SSRv, DePontieu_2022_ApJ}.

Solving the coronal heating problem involves three key steps \citep{Klimchuk_2006_SoPh}: understanding how mechanical energy is transferred to the corona, how energy dissipates within the corona, and how the corona thermally responds to heating events. 
\HK{This research aims to enhance our comprehension of the first step}{
This research aims to improve our understanding of the first step}, energy transfer. Although energy transport by flux emergence possibly plays a significant role \citep{Cranmer_2010_ApJ, Wang_2020_ApJ, Wang_2022_SolPhys}, transverse fluctuations \citep[or Alfv\'enic waves,][]{Alfven_1942_Nature, McIntosh_2011_Nature}, have been considered a promising energy carrier due to their highly efficient nature of field-aligned energy transport.
Various observations and simulations investigate the heating scenario by the transverse waves \citep[see the recent reviews by][]{VanDoorsselaere_2020_SSRv, Morton_2023_RvMPP} but its quantitative contribution is still under investigation.

\HK{}{The observations by the Coronal Multi-channel Polarimeter \citep[CoMP,][]{Tomczyk_2008_SoPh} have revealed that the power spectra of coronal transverse waves exhibit a distinct peak around $\HKK{}{4 \ \rm mHz}$. This feature is globally present in terms of both spatial and temporal characteristics \citep{Tomczyk_2007_Science, Tomczyk_2009_ApJ, Morton_2015_NatCo, Morton_2019_NatAs}. The peak frequency corresponds to the typical time scale of the solar p-modes, which are acoustic eigen oscillations in the convection zone \citep{Leighton_1962_ApJ, Ulrich_1970_ApJ, Deubner_1975_AA}. 
These results suggest that the horizontal motion of magnetic flux tubes due to granular buffeting \citep{Spruit_1981_AA, Steiner_1998_ApJ, Choudhuri_1993_SoPh, Musielak_2002_AA, Fujimura_2009_ApJ} is not only a source of coronal transverse waves but also that the p-modes may contribute to their generation by transferring a fraction of longitudinal wave energy into transverse wave energy. The transfer mechanism is still under debate, with several proposed theories. These include the direct excitation process of transverse waves by the interaction between flux tubes and longitudinal waves below the equipartition layer \citep[so called mode absorption,][]{Bogdan_1996_ApJ, Hindman_2008_ApJ, Riedl_2019_AA, Skirvin_2023_ApJ}, as well as
the mode conversion process from longitudinal to transverse waves at the equipartition layer where the local sound speed and Alfv{\'e}n speed become equal \citep{Schunker_2006_MNRAS, Cally_2007_AN, Jess_2012_ApJ, Wang_2021_ApJ,  Shimizu_2022_ApJ}. 
Additionally, another mode conversion from fast to Alfv{\'e}n waves around the refraction height of the fast waves, which is inherently a three-dimensional process unlike the mode conversion at the equipartition layer and the mode absorption \citep{Cally_2008_SoPh, Cally_2011_ApJ, Hansen_2012_ApJ, Khomenko_2012_ApJ}.}

\HK{}{Transverse oscillations of plasmas in coronal loops without obvious damping, which are called decayless oscillations, are widely used for proxies of transverse waves in the corona \citep{Wang_2012_ApJ, Tian_2012_ApJ, Nistico_2013_AA}. Recent high-cadence observations conducted by the Extreme Ultraviolet Imager \citep[EUI, ][]{Rochus_2020_AA} aboard the Solar Orbiter \citep[SolO, ][]{Muller_2020_AA} have identified the existence of high-frequency decayless oscillations (periods $< 100 \ \rm s$) in both active regions and the quiet Sun \HK{}{regions} \citep{Zhong_2022_MNRAS, Petrova_2023_ApJ, Li_2022_AA, Shrivastav_2023_aa}. Furthermore, \citet{Lim_2023_ApJ} have \HK{performed the meta-analysis of their data and}{analyzed the observed decayless oscillations and} revealed that high-frequency transverse waves play a more dominant role in coronal heating than low-frequency (periods $>100 \ \rm s$) transverse waves.}

Spicules, which are characterized as chromospheric jets embedded in the corona \citep[see the reviews by, e.g., ][]{Beckers_1968_SoPh, Beckers_1972_ARAA, Sterling_2000_SoPh, Tsiropoula_2012_SSRv, Skirvin_2023_AdSpR}, frequently exhibit transverse oscillations, used as proxies for the transverse waves \HK{as well}{} \citep[see the recent review by ][and references therein]{Jess_2023_LRSP}. 
\HK{In addition, the dissipation of transverse spicule oscillations plays a significant role in heating the surrounding \HK{solar atmosphere}{coronal plasma} \citep{Antolin_2018_ApJ}. It is worth noting that alternative mechanisms such as reconnection or \HK{ambipolar diffusion}{Joule heating} have been proposed as candidates for \HK{spicule heating}{the coronal heating by spicules} \citep{MartinezSykora_2017_Science, Samanta_2019_Science}.}{Several observations have provided evidence that high-frequency spicule oscillations, of which typical period ranging from $20$ to $50 \ \rm s$, carry substantial energy flux, contributing to both chromospheric and coronal heating \citep{Okamoto_2011_ApJ, Srivastava_2017_SciRep, Bate_2022_ApJ}. These results support the significance of high-frequency transverse waves for coronal heating.}

\HKK{By combining observations and numerical simulations, \citet{Jess_2012_ApJ} have clarified that at least a fraction of low-frequency transverse oscillations in (type-I) spicules are generated through mode conversion occurring at the equipartition layer \citep{Schunker_2006_MNRAS, Cally_2007_AN}, and have suggested that the driver of the spicule oscillations is the p-modes because of the correspondence of the period of the spicule oscillations. 
However, the exact mechanism for exciting high-frequency spicule oscillations is still under investigation. One possible explanation is presented by \citet{Shoda_2018_ApJ}, drawing from the concept of the mode conversion. This work suggests that longitudinal waves, excited by vertical granular motions or p-modes \citep{Lighthill_1952_RSPSA}, steepen into shocks and produce a cascade of the wave energy towards higher frequencies \citep{Reardon_2008_ApJ}. As these high-frequency longitudinal waves undergo the mode conversion at the equipartition layer, high-frequency transverse waves, which are responsible for high-frequency spicule oscillations, are produced. 
Nevertheless, the extent to which p-modes contribute to the generation of high-frequency spicule oscillations, as compared to vertical granular motions, remains uncertain. This is because observations are not able to distinguish their contribution in terms of oscillation period due to the effect of shock cascading.
Hence, our study aims to explore the potential role of p-modes in generating high-frequency spicule oscillations. We use a two-dimensional magneto-convection simulation to focus on the two-dimensional energy transfer system from longitudinal to transverse waves. }{High-frequency spicule oscillations cannot be generated by horizontal granular motions due to their significantly longer time scale \citep[$300$-$1000 \ \rm s$, ][]{Schrijver_1997_ApJ}. Longitudinal waves are considered one of the potential mechanisms for generating these oscillations, as demonstrated in previous numerical studies through processes such as mode conversion \citep{Shoda_2018_ApJ} or mode absorption \citep{Gao_2023_ApJ, Skirvin_2023_ApJ}. Therefore, p-modes may play a role in generating high-frequency spicule oscillations. However, the contribution of p-modes relative to granulations remains uncertain, because both p-modes and vertical granular motions generate longitudinal waves and observations cannot differentiate their contributions in terms of oscillation period. Hence, our study aims to explore the potential role of p-modes in generating high-frequency spicule oscillations using a two-dimensional magneto-convection simulation.}

\section{\HK{}{Methods}}

\subsection{\HK{}{Simulation Setup}} \label{sec:simulation_setup}

We perform a two-dimensional numerical simulation that seamlessly covers the upper part of the solar convection zone and the corona. 
To this end, we use RAMENS\footnote{RAdiation Magnetohydrodynamics Extensive Numerical Solver} code \citep{Iijima_2015_ApJ, Iijima_2016_PhD, Iijima_2017_ApJ, Wang_2021_ApJ, Kuniyoshi_2023_ApJ}, in which we solve the compressible magnetohydrodynamic equations with gravity, radiation, and thermal conduction. 
The basic equations are given in the conservation form as follows.
\begin{align}
\label{eq:mhd_eqs}
    & \frac{\partial \rho} {\partial t} + \nabla \cdot (\rho \boldsymbol{v} )  = 0, \\
    & \frac{\partial (\rho \boldsymbol{v})}{\partial t} 
      +\nabla \cdot \left[ \rho \boldsymbol{v} \boldsymbol{v}
      + \left( p+\frac{\boldsymbol{B}^2}{8\pi} \right) \boldsymbol{ \underbar I}
      - \frac{\boldsymbol{B} \boldsymbol{B}}{4 \pi}  \right]
      = \rho \boldsymbol{g}, \\
    & \frac{\partial \boldsymbol{B}}{\partial t}+\nabla \cdot (\boldsymbol{vB}-\boldsymbol{Bv})=0, \\
    & \frac{\partial e}{\partial t} 
      + \nabla \cdot \left[
     \left(e+p+\frac{\boldsymbol{B}^2}{8\pi}\right)\boldsymbol{v} -\frac{1}{4\pi} \boldsymbol{B}(\boldsymbol{v}\cdot\boldsymbol{B}) \right] \\
    & =\rho \boldsymbol{g}\cdot \boldsymbol{v}+Q_{\mathrm{cnd}}+Q_{\mathrm{rad}} \nonumber , 
\end{align}
where $\rho$ is the mass density,  $\boldsymbol{v}$ is the gas velocity, $\boldsymbol{B}$ is the magnetic field, $e=e_{\rm int} + \rho \boldsymbol{v}^2/2 + \boldsymbol{B}^2/8\pi$ is the total energy density, $e_\mathrm{int}$ is the internal energy density,
$p$ is the gas pressure, $\boldsymbol{g}$ is the gravitational acceleration, and $\boldsymbol{\underbar I}$ is unit tensor.
$Q_\mathrm{cnd}$ and $Q_\mathrm{rad}$ denote the heating by thermal conduction and radiation, respectively. 

The radiation $Q_{\rm rad}$ is determined through a combination of optically thick and thin components \HK{}{using a bridging law \citep{Iijima_2016_PhD}}.
In calculating the optically-thick radiation, the frequency-averaged (i.e., grey-approximated) radiative transfer is directly solved under the local thermodynamic equilibrium (LTE) approximation, \HK{}{using the Rosseland mean opacity obtained from the OPAL opacity \citep{Iglesias_1996_ApJ}}. 
\HK{For simplicity, the gray approximation is applied in the convection zone. }{}
The optically-thin radiation is calculated from the loss function retrieved from the CHIANTI atomic database ver. 7.1, \HK{}{assuming the coronal abundance} \citep{Dere_1997_AAS,Landi_2012_ApJ}. 
Since the loss function from CHIANTI is defined in $T \ge 10^4$ K, we employ the loss function from \citet{Goodman_2012_ApJ} in the lower-temperature range ($T \le 10^4$ K) and smoothly connect the two functions using a bridging law \cite[see][for detail]{Iijima_2016_PhD}. 
\HK{The equation of state is computed based on the LTE assumption, considering the six most abundant elements in the solar atmosphere (H, He, C, N, O, Ne).}{The equation of state is calculated under the LTE assumption, taking into account the six most abundant elements in the solar atmosphere (H, He, C, N, O, Ne). For specific details regarding the abundance and states of each element, refer to \citet{Iijima_2016_PhD}. A significant portion of the internal energy near the solar surface is derived from the latent heat generated by changes in the internal states of atoms and molecules. The balance between this latent heat and radiative transfer plays a crucial role in accurately reproducing the granulation \citep{Stein_1998_ApJ}.} The field-aligned thermal conduction of a fully-ionized plasma \citep{Spitzer_1953_PhysRev} is employed to calculate $Q_{\rm cnd}$. Although the assumption of full ionization is invalid in the chromosphere, it does not significantly influence the simulation because the thermal conduction in the chromosphere is minor. The detailed numerical procedure is found in \citet{Iijima_2016_PhD}.

Letting $x$-axis be horizontal and $z$-axis be vertical, the simulation domain covers a spatial extent of $L_x=20 \ \rm Mm$ in the $x$ direction and encompasses a vertical range from $2 \ \rm Mm$ below the surface to $14 \ \rm Mm$ above it, resulting in a total range of $L_z=16 \ \rm Mm$ in the $z$ direction. $z=0 \ \rm Mm$ is defined where the horizontally averaged optical depth is unity.
The grid size is uniformly set to $25 \ \rm km$ in $x$ direction and $25 \ \rm km$ in $z$ direction. The periodic boundary conditions are applied in $x$ direction.
We consider the loop-aligned simulation domain that extends from the upper convection zone to the top of the coronal loop, corresponding to one half of a symmetric closed loop. \HK{}{It consists of multiple flux tubes, associated with the network magnetic fields \citep{Gabriel_1976_RSPTA}.}
Following \citet{Matsumoto_2016_MNRAS},  we assume a half circle loop model for the gravitational acceleration as follows:

\begin{align}
\label{eq:gravity}
    \boldsymbol{g} = - \frac{g \cos{\theta}}{(1 + h / R_{\rm sun})^2} \hat{\boldsymbol{z}},
\end{align}

\noindent \HK{}{where $g=2.74 \times 10^4 \ \rm cm \ s^{-2}$, $R_{\rm sun}=6.96 \times 10^{10} \ \rm cm$, $\theta = z / r$, $h = r \sin{\theta}$, $r=\HK{}{2}L_z / \pi$, and $\hat{\boldsymbol{z}}$ is the unit vector in the $z$-direction.}

The \HK{}{bottom} boundary condition is open for flow, mimicking the convective energy transport from the deep convection zone (see \citet{Iijima_2016_PhD} for detail).
To ensure complete reflection of the Poynting flux at the top boundary (top of the coronal loop), we apply a reflective boundary that sets $v_z$ and $B_x$ to zero, while the other variables take on the same values as those one grid below. The reflected Poynting flux corresponds to the one injected from the other side of the loop.
To sustain the coronal temperature, we impose artificial heating at the top boundary so that the temperature $T$ is fixed to $1.0 \times 10^6$ K at the top. This artificial heating does not violate the scope of this work because our interest is in the transverse dynamics of spicules, not the amount of heating in the corona.

\begin{figure*}[!t]
  \centering
  \includegraphics[width =13cm]{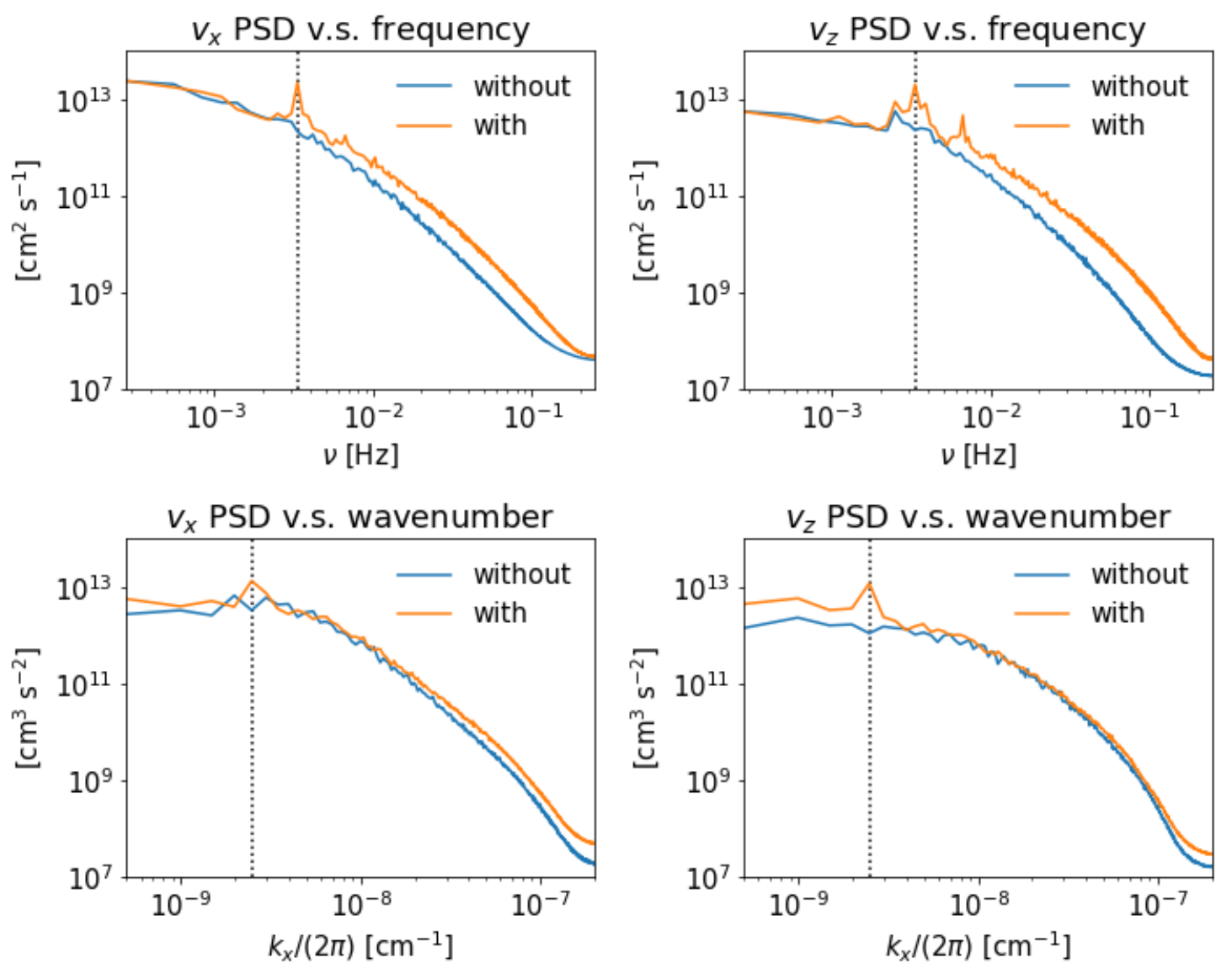}
  \caption{Power spectra of $v_x$ and $v_z$ versus the frequency and horizontal wavenumber at $z=0 \ \rm Mm$. The black dashed lines correspond to the typical frequency $\nu$ of $P_{\rm p}^{-1}=3.3 \times 10^{-3} \ \rm Hz$ (top row) and wavenumber $k_x/(2 \pi)$ of $\lambda_{\rm p}^{-1}=2.6 \times 10^{-9} \ \rm cm^{-1}$ (bottom row) of the p-modes.}
  \label{fig:psd_vx_vz}
\end{figure*}

We conduct the simulation in two stages: the first stage is performed without considering p-modes (without p-mode stage), and the second stage incorporates the p-modes (with p-mode stage). In the without p-mode stage, the initial ($t=0$) condition in the convection zone is given by Model S \citep{Christensen-Dalsgaard_1996_Science}. 
Above the surface, the initial condition is calculated by the isothermal stratification. A uniform vertical magnetic field with a strength of \HK{$5 \ \rm G$}{$10 \ \rm G$} is initially imposed. After $3 \ \mathrm{hours}$ of integration, the convection is relaxed to a quasi-steady state, in which the enthalpy flux injected from the bottom boundary nearly equals the radiative flux. 
Following this, we perform an additional one hour of integration and utilize the numerical data obtained during this period $(10,800 \ {\rm s} < t < 14,400 \ {\rm s})$. 

We use the snapshot of the without p-mode stage at $t=14,400 \ \rm s$ as the initial condition for the with p-mode stage. In this stage, we modify the bottom boundary condition to account for the influence of the p-modes because the depth of the convection zone in our simulation is not sufficient for the p-modes to develop \citep{Finley_2022_AA}. \HK{The longitudinal perturbation is excited with the vertical velocity $\delta v_z$, the density $\delta \rho$, and the pressure $\delta p$ as follows,}{We introduce longitudinal waves into the simulation by adding vertical velocity ($\delta v_z$), density ($\delta \rho$), and pressure perturbations ($\delta p$) to the local values of $v_z$, $\rho$, and $p$ in addition to the convection motions. They are defined as follows:}

\begin{align}
\label{eq:p_mode}
    & \delta v_z = v_{\rm p} \sin (2 \pi t / P_{\rm p}) \cos (2 \pi x / \lambda_{\rm p}), \\
    & \delta \rho = \rho (\delta v_z / C_s), \\
    & \delta p = \gamma p (\delta v_z / C_s),
\end{align}
where $C_s=\sqrt{\gamma p / \rho}$ is the sound speed and $\gamma = 5/3$ corresponds to the specific heat ratio of adiabatic gas. $P_{\rm p}=300 \ \rm s$ and $\lambda_{\rm p}=4.0 \ \rm Mm$, which are the typical values for the period and horizontal wavelength of p-modes obtained by previous observations \citep{Leighton_1962_ApJ, Christensen-Dalsgaard_2002_RvMP, Oba_2017_ApJ, McClure_2019_SoPh}. The longitudinal waves produced by the p-mode-like driver propagate upward and reach the photosphere. \HK{As shown in Figure \ref{fig:vz_diff},}{} Following the observation of p-modes by \citet{Oba_2017_ApJ}, $v_{\rm p}$ \HK{}{in Equation (\ref{eq:p_mode})} is a constant value ($=4.7 \times 10^4 \ \rm cm \ s^{-1}$) designed so that the power of $v_z$ at $z=0 \ \rm Mm$ in the without and the with p-mode stage correlate as 

\HK{\begin{align}
 \langle v_{z, \rm with}^2 \rangle_{t,x} - \langle v_{z, \rm without}^2 \rangle_{t,x}
 = \langle v_{z, \rm without}^2 \rangle_{t,x},
\end{align}}{\begin{align}
\label{eq:pmode_rms}
 \frac{\langle v_{z, \rm with}^2 \rangle}{\langle v_{z, \rm without}^2 \rangle}=2 ,
\end{align}}

\noindent where the operator $\langle \rangle$ denotes the temporal and horizontal averaging. \HK{}{While \citet{Oba_2017_ApJ} did not observe the horizontal velocity field, for reference, we present the ratio of the power of $v_x$ at $z=0 \ \rm Mm$ as follows,}

\begin{align}
\label{eq:pmode_rms_x}
 \HK{}{\frac{\langle v_{x, \rm with}^2 \rangle}{\langle v_{x, \rm without}^2 \rangle}=1.3,} 
\end{align}

\noindent \HKK{The simulation was run for $4 \ \rm hour$, with the last hour specifically dedicated to analyzing the numerical data.}{It is worth noting that $v_x$ is amplified due to the p-mode-like driver despite the driver not being applied to $v_x$ because the longitudinal waves generated by the driver propagate isotropically through the convection zone to the photosphere. The simulation was run for $4 \ {\rm hours} \ (14,400 \ {\rm s} < t < 28,800 \ {\rm s})$, with the last hour $(25,200 \ {\rm s} < t < 28,800 \ {\rm s})$ specifically dedicated to analyzing the numerical data. It is worth noting that the temporal averages of Equation (\ref{eq:pmode_rms}) and (\ref{eq:pmode_rms_x}) are calculated over the duration $25,200 \ {\rm s} < t < 28,800 \ {\rm s}$.}

\begin{figure*}[!t]
  \centering
  \includegraphics[width = 15 cm]{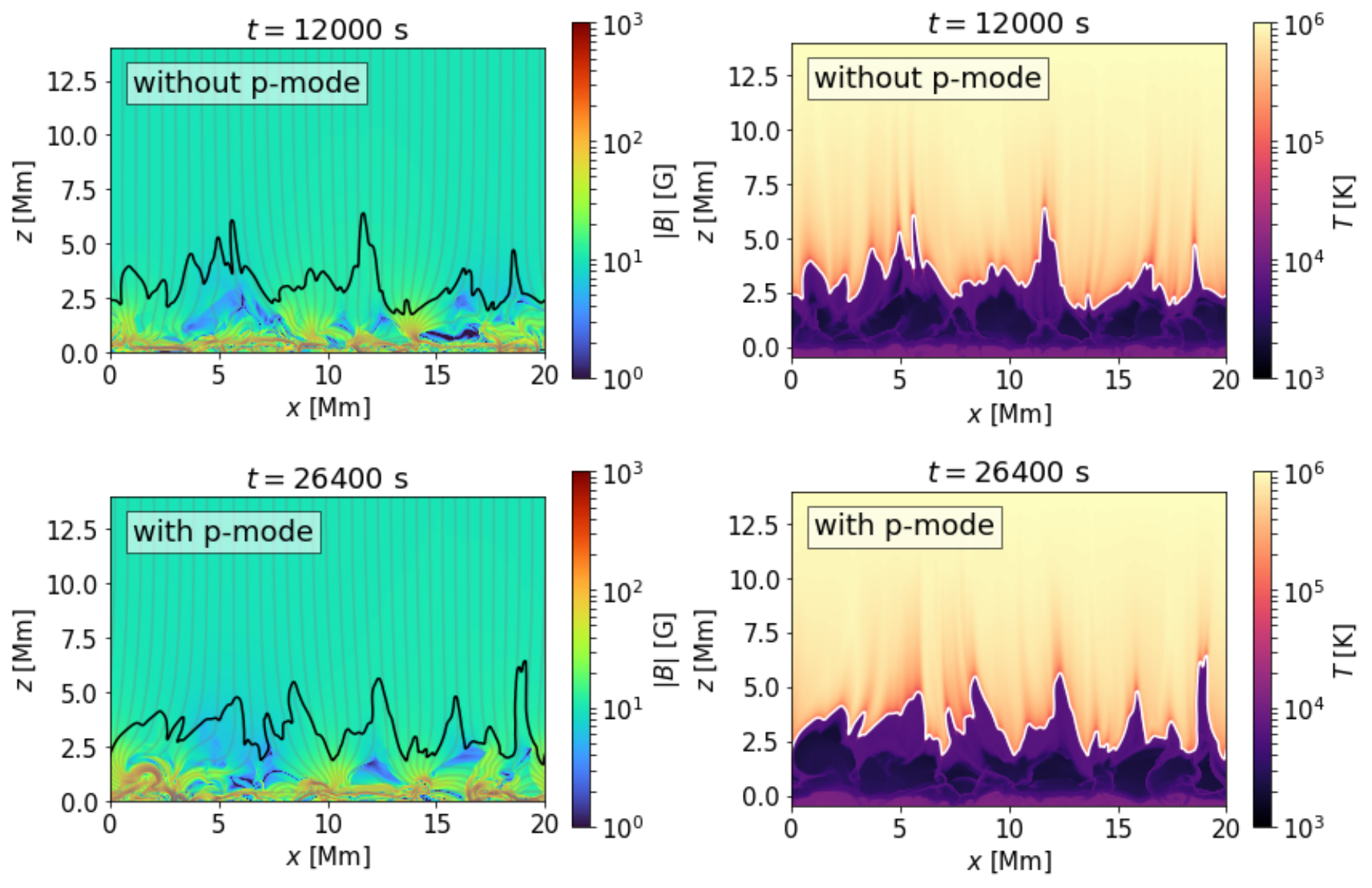}
  \caption{The snapshots consist of the (left column) magnetic field strength maps in color with magnetic field lines in grey lines and the transition region defined by $T=10^4 \ \rm K$ highlighted by black lines, and (right column) temperature maps in color with the transition region highlighted by white lines. Both columns include cases for both the without and with p-mode stages. \HKK{}{An animation of this figure is available that shows the temporal evolutions of the magnetic field and temperature distributions and the positions of the transition region over a period of $1,800 \ \rm s$}.}
  \label{fig:te_snapshot}
\end{figure*}

Figure \ref{fig:psd_vx_vz} illustrates the velocity power spectra, denoted as $E_j^{\rm fq}$ and $E_j^{\rm wn}$, as a function of frequency and wavenumber at $z=0 \ \rm Mm$. Here, $j=x,z$, and these quantities are defined as follows:
\begin{align}
\label{eq:psd_vx_vz}
    & E_j^{\rm fq} (\omega) = \frac{1}{\Delta \omega} \left| \sum_{n_t=1}^{N_t} \HKK{}{v_j(n_t, n_x)} e^{-2 \pi i \omega n_t / N_t} \right|^2, \\
    & E_j^{\rm wn} (k_x) = \frac{1}{\Delta k_x} \left| \sum_{n_x=1}^{N_x} \HKK{}{v_j(n_t, n_x)} e^{-2 \pi i k_x n_x / N_x} \right|^2,
\end{align}
\HK{where $\nu$ is the frequency and $k_x$ is the horizontal wavenumber.}{where the indices for the time ($t$) and horizontal ($x$) directions are denoted as $n_t$ and $n_x$, respectively, with lengths $N_t$ and $N_x$. We sampled frequencies ($\nu = \omega / 2 \pi$) and horizontal wavenumbers ($k_x$) in increments of \HKK{}{$\Delta \omega = 2\pi / (N_t \Delta t)=1.7 \times 10^{-3} \ \rm s^{-1}$} and \HKK{}{$\Delta k_x = 2\pi / (N_x \Delta x)= 3.1 \times 10^{-9} \ \rm cm^{-1}$}, where $\Delta x$ represents the grid size in the $x$ direction (which is $25 \ \rm km$), and $\Delta t$ is equal to $2 \ \rm s$.}
\HK{}{The power spectra exhibit sensitivity to the bottom boundary condition (i.e., with or without p-mode stages).}
The power spectra for the with p-mode stage exhibit distinct peaks at locations that correspond to the typical values of the p-modes, $\nu = P_{\rm p}^{-1}$ and $k_x = 2 \pi \lambda_{\rm p}^{-1}$. 



 \HK{The simulation was run for $4 \ \rm hour$, with the last hour $(25200 \ \rm s < t < 28800 \ \rm s)$ specifically dedicated to analyzing the numerical data.}{}

\subsection{\HK{}{Spicule Oscillation Analysis Method}}
\label{sec:sp_oscil}

Figure \ref{fig:te_snapshot} displays the snapshots of the magnetic field and temperature distribution in both the without and with p-mode stages. After the relaxation of the convection, expanding magnetic flux tubes are formed of which footpoints are concentrated into kilogauss magnetic fields. 
\HKK{It reveals the presence of multiple chromospheric plasmas that exhibit spicule-like structures.}{In addition, Figure \ref{fig:te_snapshot} reveals several spicule-like structures consisting of plasmas with chromospheric temperatures ($T \sim 10^4 \ \rm K$).} We utilize a fully automatic method with three steps to detect spicules and measure the period and velocity amplitude of the transverse oscillations in them. The method consists of the following steps:

\begin{enumerate}
  \item Identifying the spicules within the simulated data.
  \item Tracking the transverse oscillations of the identified spicules at a specific height.
  \item Deriving periods and velocity amplitudes of the spicule oscillations.
\end{enumerate}

\noindent In constructing our three-step method, we draw upon the NUWT\footnote{Northumbria University Wave Tracking} code as a reference \citep{Morton_2013_ApJ, Thurgood_2014_ApJ, Weberg_2018_ApJ}. 

\begin{figure}[!t]
  \centering
  \includegraphics[width = 8cm]{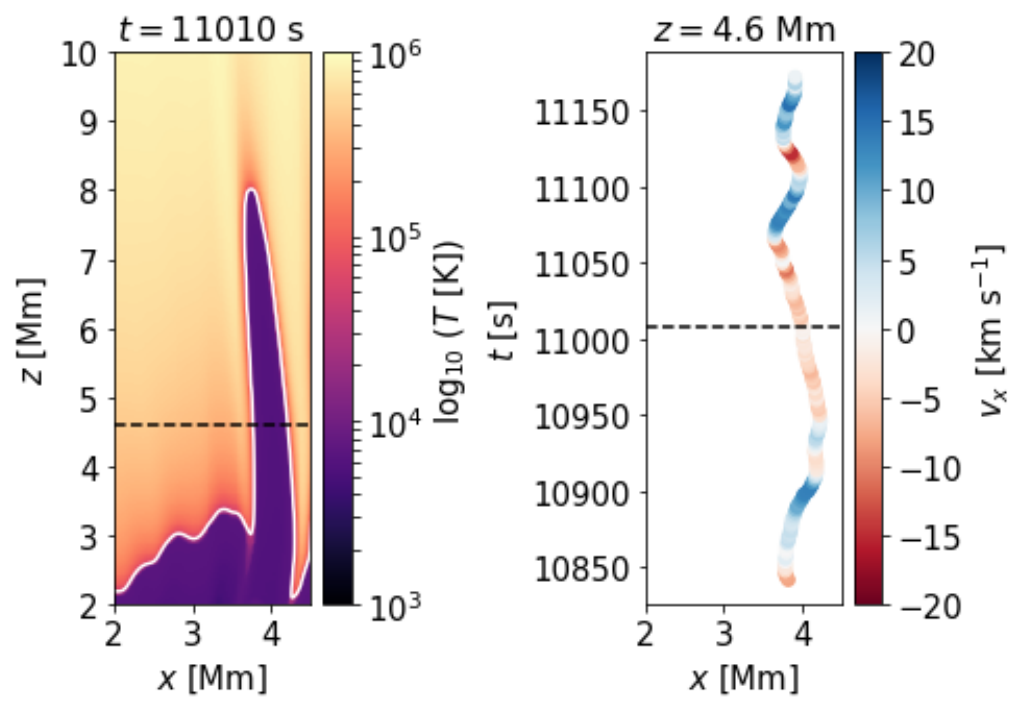}
  \caption{(Left) A zoomed-in view of the temperature map, specifically focusing on a spicule. The white line indicates the transition region. The black dashed line shows the height corresponding to the right panel ($z=4.5 \ \rm Mm$). (Right) \HKK{}{The mean horizontal velocity $v_x$ within a spicule is represented by the color of the dots, while the position of each dot corresponds to the mean horizontal coordinate ($x$) between the right and left edges during the lifetime of the spicule. The black dashed line shows the time corresponding to the left panel ($t=11,010 \ \rm s$).}}
  \label{fig:fitting_description}
\end{figure}

\emph{Step 1}: 
\HK{In our simulation, we employ an automated approach to identify spicules by tracking the temporal and horizontal evolution of the top of each spicule. At the edges of the spicules, i.e., the transition region, the temperature is measured to be $T=40,000 \ \rm K$. The spicule top is determined as the local maximum in the distribution of the transition region height across both the temporal ($t$) and horizontal ($x$) directions. Note that only local maxima with a distance between adjacent local minima exceeding $2 \ \rm Mm$ are classified as spicules. In the left panel of Figure \ref{fig:fitting_description}, a temperature map is displayed, emphasizing a spicule where both the edges and center are highlighted.}{We utilize an automated approach to identify spicules, considering them as local minima of temperature below $40,000 \ \rm K$ across both temporal ($t$) and horizontal ($x$) dimensions at a specific height. Note that only local minima with widths in the $x$-direction less than $1,100 \ \rm km$ are categorized as spicules. This criterion aligns with the maximum spicule width \HK{}{observed} in the quiet Sun atmosphere \citep{Pasachoff_2009_SoPh}. \HKK{}{The lifetime of each spicule is determined by identifying the start and end times of the local minimum.} In the left panel of Figure \ref{fig:fitting_description}, a temperature map is displayed, emphasizing \HK{a spicule where both the edges and center are highlighted.}{the edges of the spicule by white lines.}}

\emph{Step 2}: 
\HK{For a specific height, we define the center of the right and left edges of the spicule as the spicule center, and the distance between these edges as the spicule width. If, at a specific height located within $2 \ \rm Mm$ below the spicule top, the spicule width measures $700 \ \rm km$, we track the displacement of the spicule center at that height. However, if the spicule width is smaller than $700 \ \rm km$ within $2 \ \rm Mm$ below the spicule top, we instead track the displacement of the spicule center at a height $2 \ \rm Mm$ below the spicule top. Note that the value $700 \ \rm km$ is the maximum spicule width observed by \citet{DePontieu_2007_ApJ}. We record the temporal variation of the horizontal velocity $v_x$ along the displacement, considering it as the transverse velocity. This assumption is based on the nearly vertical orientation of the spicules as waveguides.}{
At the given height and time, we calculate the mean horizontal ($v_{x}$) and perpendicular velocities (to the local magnetic field, $v_{\perp}$) between the right and left edges of the spicules. We then proceed to record the temporal evolution of them.}
In the right panel of Figure \ref{fig:fitting_description}, we present an example of the recorded \HKKK{$v_{\perp}$}{} $v_x$ along the displacement, demonstrating an oscillatory pattern.

\emph{Step 3}:
To get the oscillation period, we utilize the fast Fourier transform (FFT) on the recorded temporal variation and generate the power spectra, denoted as $E^{\rm fq}_{\rm sp}$, with respect to frequency $\nu$ defined as,


\begin{align}
\label{eq:fft_vel}
E_{k, \rm sp}^{\rm fq} (\omega) = \frac{1}{\Delta \omega} \left| \sum_{n_t=1}^{N_{k, \rm sp}} v_{k, \rm sp}(n_t) e^{-2 \pi i \omega n_t / N_{k,\rm sp}} \right|^2
\end{align}

\noindent where \HKK{}{$k=x, \perp$, $v_{k, \rm sp}$ denotes the recorded velocity of the spicule during its lifetime, $N_{k,\rm sp}$ represents the sample size of $v_{k, \rm sp}$ for each spicule.} 
We select the frequency to be that of the significant wave component\HK{when the ratio $R_v(\nu)$, defined as given below, satisfies the following equation:}{, fulfilling the following \HK{equation}{condition}:}


\begin{align}
\label{eq:selection}
\frac{E^{k, \rm fq}_{\rm sp}(\omega) \HK{}{\Delta \omega}}{ \sum_{\omega} E^{k, \rm fq}_{\rm sp}(\omega) \Delta \omega } > 0.05.
\end{align}

\noindent Note that the frequency components which show at least $3/4$ of an oscillatory cycle are chosen. 
This approach leads to the selective detection of the high-frequency oscillations because it cannot identify oscillations with periods longer than the typical lifetime of spicules, which is on the order of a few minutes.
The individual oscillation often exhibits multiple\RJM{signatures of the}{} significant frequency components. We fit the time evolution of the horizontal velocity oscillations with a sinusoidal and linear function $v_{\rm fit} (t)$ as

\begin{align}
\label{eq:fitting_vel}
v_{\rm fit} (t) = \sum_i^{N_{\rm mode}} \left[ A_{v,i} \sin \left( \frac{2\pi t}{P_i} + \phi_i \right) \right] + c_1 t + c_2,
\end{align}

\noindent where the subscript $i$ expresses the individual wave component of the superposition, $N_{\rm mode}$ represents the number of oscillation modes chosen according to Equation (\ref{eq:fitting_vel}), $A_{v,i}$ is the velocity amplitude, $P_i$ is the period (inverse of the selected frequency). We obtain $A_{v,i}$, $\phi_i$, $c_{1}$, and $c_{2}$ through the fitting procedure.

\section{Analysis and Results}
\label{sec:analysis_results}

\subsection{\HK{}{Energy Transfer between Waves}}
\label{sec:energy_transport_waves}

\begin{figure}[!t]
  \centering
  \includegraphics[width = 8 cm]{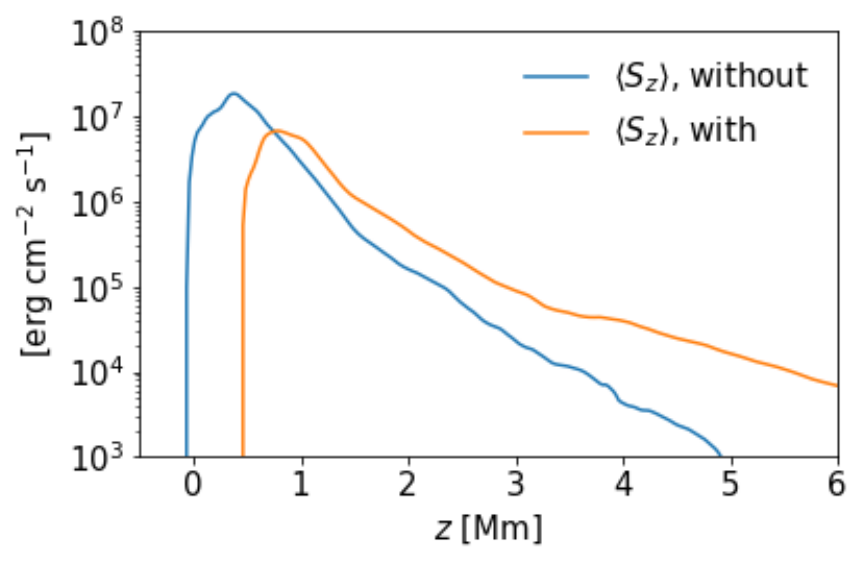}
  \caption{\MS{$\langle S_z \rangle$}{The height dependence of the averaged vertical Poynting flux (denoted as $\langle S_z \rangle$)} during the without and with p-mode stage. \HK{Shaded areas represent the 1$\sigma$ sections of the equipartition layer $H_{\rm eq}$ for both stages, with significant overlap between them.}{}}
  \label{fig:fmg_fac}
\end{figure}

Figure \ref{fig:fmg_fac} illustrates the distributions of the \HK{$\langle F_{\rm ac, z} \rangle$ and $\langle F_{\rm mg, z} \rangle$}{$\langle S_z \rangle$} for the stages without and with p-modes, where  

\begin{align}
\label{eq:s_z}
S_z = B_x^2 v_z / 4 \pi - v_x B_x B_z / 4 \pi.
\end{align}

\noindent \HKK{}{$\langle S_z \rangle$ in the with p-mode stage is enhanced above $z=1 \ \rm Mm$ compared to that in the without p-mode stage. In contrast, the increase in energy flux transported by longitudinal waves in the with p-mode stage initiates below $z = 0 \ \rm Mm$ as a consequence of the p-mode-like driver. (see Section \ref{sec:simulation_setup}). Therefore, the increase in $\langle S_z \rangle$ signifies the transfer of energy from the longitudinal waves triggered by the driver to the transverse waves. It's worth noting that $\langle S_z \rangle$ in the with p-mode stage below z=0.5 Mm is negative because of the greater amplification of $\langle S_z^{\rm emerge} \rangle = B_x^2 v_z / 4 \pi \ (<0)$ compared to $\langle S_z^{\rm shear} \rangle = - v_x B_x B_z / 4 \pi \ (>0)$. Due to the temporal and horizontal averaging applied to the energy fluxes, the amplification of $\langle S_z \rangle$ likely involves multiple mechanisms, such as the mode conversion or mode absorption, which are local phenomena.}

\subsection{\HK{}{Statistical Properties of Spicule Oscillations}}
\label{sec:sp_oscill}

\begin{figure}[!t]
  \centering
  \includegraphics[width = 8cm]{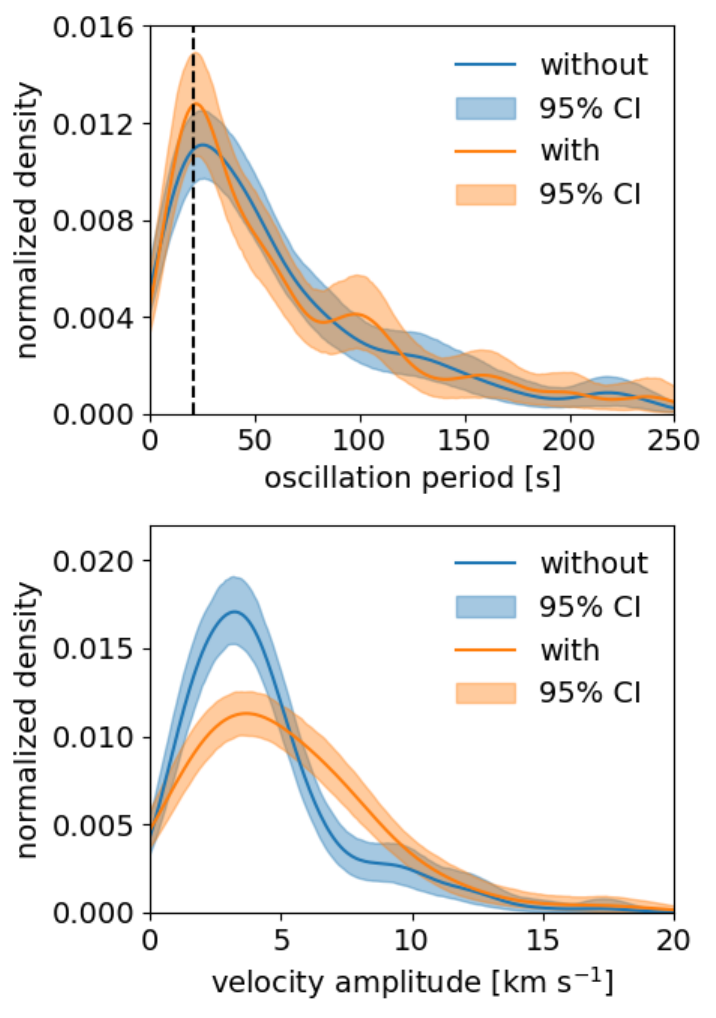}
  \caption{The kernel density estimation for the distributions of (top) the oscillation periods $P_v$ and (bottom) the velocity amplitude $A_v$ derived from perpendicular velocities $v_{\perp}$. Each panel displays parameters in the without (blue line) and with p-mode stage (orange line) with a $95\%$ confidence interval (CI) for the estimates calculated using bootstrapping. The black dotted line in the top panel shows $\langle \tau_{\rm eq} \rangle$.}
  \label{fig:hist_vel_prd}
\end{figure}

\begin{table*}[t]
\centering
\caption{Spicule oscillation properties in the without and with p-mode stage derived from perpendicular velocities $v_{\perp}$}.
\begin{tabular}{ccccccc}
\hline \hline 
\multicolumn{1}{c}{} & \multicolumn{3}{c}{without (mean $\pm$ standard error)} & \multicolumn{3}{c}{with (mean $\pm$ standard error)} \\
\cmidrule(rl){2-4} \cmidrule(rl){5-7} 
{height} & {$N_{\rm os, without}$} & {$\bar{P}_{v, \rm without}$} & {$\bar{A}_{v, \rm without}$} & {$N_{\rm os, with}$} & {$\bar{P}_{v, \rm with}$} & {$\bar{A}_{v, \rm with}$} \\
$(\rm km)$ & {} & {$(\rm s)$} & {$(\rm km \ s^{-1})$} & {} & {$(\rm s)$} & {$(\rm km \ s^{-1})$} \\
\hline
$4200$ & $220$ & $55 \pm 4$ & $4.4 \pm 0.2$ & $140$ & $65 \pm 5$ & $5.9 \pm 0.4$ \\
$4400$ & $215$ & $63 \pm 4$ & $4.6 \pm 0.2$ & $148$ & $60 \pm 5$ & $5.3 \pm 0.5$ \\
$4600$ & $198$ & $62 \pm 4$ & $4.1 \pm 0.2$ & $147$ & $69 \pm 5$ & $5.6 \pm 0.4$ \\
$4800$ & $173$ & $65 \pm 5$ & $4.1 \pm 0.2$ & $147$ & $69 \pm 5$ & $5.5 \pm 0.4$ \\
$5000$ & $132$ & $70 \pm 6$ & $4.5 \pm 0.3$ & $149$ & $68 \pm 5$ & $5.3 \pm 0.3$ \\
$5200$ & $123$ & $64 \pm 6$ & $3.9 \pm 0.3$ & $135$ & $69 \pm 6$ & $5.2 \pm 0.3$ \\
$5400$ & $106$ & $68 \pm 6$ & $4.1 \pm 0.3$ & $133$ & $67 \pm 6$ & $4.7 \pm 0.4$ \\
\hline
\end{tabular}
\label{tab:oscil_tab}
\end{table*}

\HK{Applying \HK{the}{our} spicule detection and oscillation fitting method \HK{described in detail in Section \ref{sec:spicule_oscillation}}{}, we have identified a total of 447 oscillations in the without p-mode stage and 407 oscillations in the with p-mode stage. These numbers are larger than those of the identified spicules because multiple oscillations are detected within many individual spicules.}{In this section, we investigate whether high-frequency spicule oscillations exhibit any signatures of energy transfer resulting from the longitudinal waves initiated by the p-mode-like driver, as shown in Section \ref{sec:energy_transport_waves}.
We have applied our spicule oscillation analysis method at various heights, focusing on regions where the number of oscillations ($N_{\rm os}$) is at least 100. This analysis encompasses heights up to $5,400 \ \rm km$, where $N_{\rm os} \geq 100$ in both the without and with stages, and the lowest height considered corresponds to the lower limit of spicule height observed in the quiet Sun \citep[$4,200 \ \rm km$,][]{Pasachoff_2009_SoPh}.} 
\HKKK{The distributions of velocity amplitude $A_v$ and oscillation period $P_v$ obtained from the horizontal velocities $v_x$ and perpendicular velocities $v_{\perp}$ within spicules are almost identical because the spicules are almost vertical due to the top boundary condition. Hence, we present the results obtained only from $v_x$ in the following section.}{}
As an illustrative example, Figure \ref{fig:hist_vel_prd} displays the density distributions of velocity amplitudes $A_v$ and oscillation periods $P_v$ at $z=4.5 \ \rm Mm$ obtained from $v_{\perp}$ within the spicules. These distributions are estimated using kernel density estimation (KDE) with a Gaussian kernel. The bandwidth parameter is selected through cross-validation\footnote{Conducted using Scikit-learn \citep{Pedregosa_2011}}, following the same method as in \citet{Morton_2021b_ApJ}. The distributions of $P_v$ reveal a dominance of high-frequency (periods $< 100 \ \rm s$) components in both stages with a peak around $20 \ \rm s$. Moreover, the distributions of $P_v$ demonstrate minimal difference between the two stages, while that of $A_v$ indicates an enhancement in the with p-mode stage compared to the without p-mode stage. These are common characteristics across all considered heights.
The distributions of $A_v$ and $P_v$ obtained from $v_{x}$ and $v_{\perp}$ within the spicules exhibit a striking degree of similarity (see the distributions derived from $v_{x}$ as depicted in Figure \ref{fig:hist_vel_prd_perp} in the Appendix). This consistency arises primarily because a lot of spicules in our simulation are vertical due to our top boundary condition, which enforces a vertical magnetic field, i.e., $B_x=0 \ \rm G$, although some of them do exhibit clear inclinations.





\HKK{}{At each of the heights investigated, we have computed the mean values of the oscillation period $\bar{P}_v$ and velocity amplitude $\bar{A}_v$.
Table \ref{tab:oscil_tab} summarizes the oscillation parameters ($N_{\rm os}$, $\bar{P}_v$, and $\bar{A}_v$) derived from perpendicular velocities $v_{\perp}$ within the detected spicules. Across different altitudes, there is no significant variation in mean parameters, and this remains within the range of standard errors. It is worth noting that the mean parameters derived from $v_{x}$ closely resemble those obtained from $v_{\perp}$ (refer to Table \ref{tab:oscil_tab_perp} in the Appendix).
Furthermore, to assess the impact of the p-mode-like driver on the mean velocity amplitude, we have calculated the ratio of the mean velocity amplitude between the with and without p-mode stage, denoted as $R_{v, \rm sp} = \bar{A}_{v, \rm with} / \bar{A}_{v, \rm without}$.  
The presence of the driver leads to an intensification of the mean velocity amplitudes by \HKK{$10$--$40 \%$ (i.e., $R_{v, \rm sp}=1.1$--$1.4$)}{$10$--$30\%$ (i.e., $R_{v, \rm sp}=1.1$--$1.3$)} when using $\bar{A}_v$ from $v_{\perp}$ and $10$--$40\%$ (i.e., $R_{v, \rm sp}=1.1$--$1.4$) when using $\bar{A}_v$ from $v_{x}$. When we assess the increase in the power of velocity amplitude $R_{v, \rm sp}^2$, we obtain $R_{v, \rm sp}^2 = 1.3$--$1.8$ when using $v_{\perp}$ and $R_{v, \rm sp}^2 = 1.2$--$1.9$ when using $v_{x}$. This intensification exceeds the power of horizontal velocity at $z=0 \ \rm Mm$ ($30\%$, see Equation (\ref{eq:pmode_rms_x})). Hence, this result is unlikely to be solely attributed to the increase in horizontal velocity caused by the p-mode-like driver. Instead, it is more likely due to the energy transfer from the longitudinal wave energy generated by the driver to transverse wave energy, as explained in Section \ref{sec:energy_transport_waves}.}

\section{Discussion} \label{sec:discussion}

\HK{The velocity amplitude and oscillation period of the transverse spicule oscillations show almost no difference, regardless of whether the p-mode-like driver is present or not (see Figure \ref{fig:hist_vel_prd}). 
This result suggests that the contribution of the p-mode to the generation of high-frequency spicule oscillations through longitudinal to transverse mode conversion is minimal.
Instead, the oscillations may be directly generated by the horizontal flows within intergranular lanes, or through the mode conversion facilitated by the shocks generated by granular upflows.}{The velocity amplitude of high-frequency spicule oscillations is enhanced by the presence of the p-mode-like driver. 
Therefore, despite accounting for the refraction of transverse (fast magnetic) waves, p-modes still have the potential to contribute to the production of high-frequency spicule oscillations, although the exact mechanism of the energy transfer cannot be definitively determined.}

\HKK{}{The distributions of oscillation periods for both stages exhibit a peak around $20 \ \rm s$. This value is consistent with the mean crossing time of acoustic waves across the equipartition layer, denoted as $\langle \tau_{\rm eq} \rangle$, which is expressed as follows \citep{Shoda_2018_ApJ}:}

\begin{align}
\label{eq:ta_sp}
\tau_{\rm eq} = \frac{1}{C_s} \left[ \frac{d}{dz} \left( \frac{v_A^2}{C_s^2} \right) \right]^{-1} \Bigg|_{C_s = v_A}.
\end{align}

\noindent \HKK{}{$\langle \tau_{\rm eq} \rangle = 21 \ \rm s$ in the without and with p-mode stage, as indicated by the dashed lines in the top panels of Figure \ref{fig:hist_vel_prd}. This result can be explained as follows: pulse-like amplifications in the velocity amplitude, lasting for approximately $20 \ \rm s$, are generated through the mode conversion at the equipartition layer \citep{Schunker_2006_MNRAS, Cally_2007_AN}, as shown in \citet{Shoda_2018_ApJ}. This finding is a consistent quantitative feature across all the heights we examined. However, it is important to note that this factor does not conclusively rule out the presence of other energy transfer systems, such as the mode absorption \citep{Bogdan_1996_ApJ, Hindman_2008_ApJ, Riedl_2019_AA, Skirvin_2023_ApJ}.}

\HK{}{Many observations employ the velocity amplitude of spicule oscillations to estimate the carried energy flux $F$ \citep{VanDoorsselaere_2014_ApJ, Jess_2023_LRSP}, using the following formula:}

\begin{align}
\label{eq:f_sp}
F = \rho v_A v_{\perp}^2.
\end{align}

\noindent $v_{\perp}$ is obtained from the mean velocity amplitude of spicule transverse oscillations observed, while $\rho$ and $v_A$ are derived from various models \citep[e.g., a bright network chromospheric model by][]{Vernazza_1981_ApJS}. Therefore, the squared value of the velocity amplitude serves as a primary indicator of the energy flux propagating within spicules. 
When we assume that all the obtained values of $\bar{A}_{v, \rm without}$ and $\bar{A}_{v, \rm with}$ represent the velocity amplitudes of propagating waves, the energy flux carried by them through spicules is amplified by $30$--$80 \%$ (i.e., $R_{v, \rm sp}^2=1.3$--$1.8$).

To analyze the nature of the detected oscillations, i.e., whether they are propagating or standing waves, we calculate their phase velocity $v_{\rm ph}$ as follows:

\begin{align}
\label{eq:v_ph}
v_{\rm ph} = \frac{2 \pi d}{P_v \Delta \phi},
\end{align}

 \noindent where $\Delta \phi$ represents the phase lag of the individual waves $\phi$ at two different heights ($5,400 \ \rm km$ and $4,800 \ \rm km$), as obtained from Equation (\ref{eq:fitting_vel}), and $d$ represents the height difference, i.e., $d=600 \ \rm km$. $\phi$ and $P_v$ are derived from $v_{\perp}$ within the spicules.  
 To detect features of the same wave at different heights, we derive the properties of individual waves at a specific height and search for similar ones at an adjacent height, following the criteria of \citet{Bate_2022_ApJ}. The considered properties (and criteria) include: (i) the $x$-position of the spicule averaged over its lifetime (with a difference not exceeding $250 \ \rm km$, i.e., $10$ grids), (ii) the lifetime of the spicule (ensuring the duration of its existence between the considered heights overlaps), (iii) the duration of the oscillation (with a difference not exceeding $50\%$), and (iv) the oscillation period (with a difference not exceeding $10\%$). Based on these criteria, we have identified 13 upwardly propagating, 7 downwardly propagating, and 6 standing waves in the without p-mode stage. In the with p-mode stage, we have identified 6 upwardly propagating, 2 downwardly propagating, and 10 standing waves.
 Waves with a phase speed $v_{\rm ph} > 500 \ \rm km \ s^{-1}$ are classified as standing waves, following the same criteria applied by \citet{Okamoto_2011_ApJ}. 
  It should be noted that the number of the identified waves is much smaller than the total number of the detected oscillations $N_{\rm os}$. This is because the transverse waves in our two-dimensional simulation are fast mode waves, which propagate not only along but also across the magnetic field lines inside the spicules.
 Considering the obtained values of $\bar{A}_{v, \rm without}$ and $\bar{A}_{v, \rm with}$ in our simulation (averaged between those at $z=5,400 \ \rm km$ and $z=4,800 \ \rm km$) presented in Table \ref{tab:phase_analy}, the energy flux propagating through spicules is intensified by around $150 \%$ ($R_{v, \rm sp}^2=2.5 \pm 1.4$) for the upward waves and $250 \%$ for the downward waves ($R_{v, \rm sp}^2=3.5 \pm 1.9$). On the other hand, there is no clear increase for standing waves (i.e., $R_{v, \rm sp}^2 = 0.9 \pm 0.8$). Thus, propagating waves within spicules are significantly enhanced by the p-mode-like driver, while standing waves are not affected. However, care must be taken in the classification of the standing waves. Upwardly and downwardly propagating waves are superposed in the temporal evolutions of the spicule oscillations obtained by our method, potentially leading to misidentification of standing waves. The more sophisticated classification, achieved by filtering upwardly and downwardly propagating waves using spatiotemporal Fourier analysis \citep{Tomczyk_2009_ApJ}, may change the result.

 \begin{table}[!t]
\centering
\caption{Velocity amplitudes and phase velocities of the propagating or standing waves in the without and with p-mode stage derived from perpendicular velocities $v_{\perp}$ within the spicules.}
\begin{tabular}{lccc}
\hline \hline 
\multicolumn{1}{c}{} & \multicolumn{1}{c}{upward $\bar{A}_{v}$} & \multicolumn{1}{c}{downward $\bar{A}_{v}$} & \multicolumn{1}{c}{standing $\bar{A}_{v}$} \\
\cmidrule(rl){2-2} \cmidrule(rl){3-3} \cmidrule(rl){4-4} 
{driver} & {$(\rm km \ s^{-1})$} & {$(\rm km \ s^{-1})$} &  {$(\rm km \ s^{-1})$} \\
\hline
without & $3.7 \pm 0.5$ & $3.7 \pm 0.4$ & $4.0 \pm 0.9$ \\
with & $5.9 \pm 0.8$ & $6.9 \pm 1.2$ & $3.8 \pm 0.8$ \\
\hline
\end{tabular}
\label{tab:phase_analy}
\end{table}


The oscillation periods obtained from horizontal velocities in simulated spicules (Figure \ref{fig:hist_vel_prd}) fall within the observed range of high-frequency spicule oscillations, as indicated by previous studies \citep{Okamoto_2011_ApJ, Bate_2022_ApJ}.
\HKK{The mean velocity amplitudes of the simulated spicule oscillations correspond well to the velocity amplitudes observed in the quiet Sun $2$--$3 \ \rm km \ s^{-1}$ \citep{Yoshida_2019_ApJ} and in the coronal hole $7.4 \ \rm km \ s^{-1}$ \citep{Okamoto_2011_ApJ}.
While they are lower than those observed in active regions $21 \ \rm km \ s^{-1}$, they remain within the reported range \citep{Bate_2022_ApJ}.}{The velocity amplitudes of the simulated spicule oscillations (Table \ref{tab:oscil_tab}) are smaller than many observed values \citep[median $\sim 10 \ \rm km \ s^{-1}$,][]{Okamoto_2011_ApJ, Pereira_2012_ApJ, Bate_2022_ApJ}, while a few studies show good correspondence \citep{Jafarzadeh_2017_ApJS, Yoshida_2019_ApJ}. This discrepancy is likely to come from the top boundary condition in our simulation. In our simulation, the top boundary condition is implemented such that the magnetic field becomes vertical, i.e., $B_x = 0 \ \rm G$, which consequently suppresses the transverse velocities of spicules.}

\HK{}{The top boundary condition ($B_x = 0 \ \rm G$) also results in the reflection of transverse waves. To compare the energy flux carried by the reflected wave at the top boundary and the incident wave at the transition region, we have estimated them by using Equation (\ref{eq:f_sp}). 
We verify that $\langle F \rangle$ at the upper boundary is $5 \%$ of that at the transition region in the absence of the p-mode-like driver, and $3 \%$ in the presence of the driver. Hence, we can disregard the influence of the top boundary on the transverse spicule oscillations (i.e., below the transition region). However, the wave modulation in the corona arising from this reflection is not to be disregarded. Consequently, our numerical model does not have the capacity to investigate the generation of the $4 \ \rm mHz$ peak observed in the coronal power spectra, as demonstrated in CoMP \citep{Tomczyk_2007_Science, Tomczyk_2009_ApJ, Morton_2015_NatCo, Morton_2019_NatAs}.}

It is important to note that our two-dimensional simulation does not account for the three-dimensional effects such as the other mode conversion, responsible for generating Alfv{\'e}n waves from fast waves \citep{Cally_2008_SoPh, Cally_2011_ApJ, Hansen_2012_ApJ, Khomenko_2012_ApJ}. In addition, spicules display three-dimensional motions, encompassing not only bulk transverse (kink-mode-like) oscillations but also rotational and cross-sectional oscillations \citep{Sharma_2017_ApJ, Sharma_2018_ApJ}. These additional factors may indeed have an impact on the properties of high-frequency spicule oscillations produced by the p-modes. Further investigations using three-dimensional simulations are necessary to explore these effects in detail.

\section{Conclusion} \label{sec:conclusion}

Our study focuses on examining the potential role of solar p-modes in generating high-frequency transverse spicule oscillations. To investigate this, we utilize a two-dimensional magneto-convection simulation that encompasses the upper convection zone to the corona. We introduce a driver in the middle of our simulation duration to generate longitudinal waves with periods and wavelengths resembling the typical values of p-modes. 

We then proceed to analyze and compare the spicule oscillation characteristics between the stages without and with the p-mode-like driver. We find that the mean velocity amplitude increases by $10$--$30 \%$ by the driver, leading to the estimated enhancement of the energy flux in the spicules by $30$–$80 \%$.
This result implies that both p-modes and granulations play a substantial role in generating high-frequency spicule oscillations. Further investigations utilizing three-dimensional simulations are necessary to explore the contribution of p-modes to producing three-dimensional spicule transverse oscillations.


\

We would like to convey our sincere appreciation to the anonymous referee for providing valuable feedback.
Numerical computations were carried out on the Cray XC50 at the Center for Computational Astrophysics (CfCA), National Astronomical Observatory of Japan.
M.S. is supported by JSPS KAKENHI Grant Number JP22K14077.
R.J.M. is supported by a UKRI Future Leader Fellowship
(RiPSAW MR/T019891/1). H. K. is also grateful for travel support provided by the UKRI Future Leader Fellowship
(RiPSAW MR/T019891/1). T.Y. is supported by the JSPS KAKENHI Grant Number JP21H01124, JP20KK0072, and JP21H04492.
This work was supported by NAOJ Research Coordination Committee, NINS, Grant Number NAOJ-RCC-2301-0301. 

%




\appendix
\section{Additional Figures and Tables}

\begin{figure}[!t]
  \centering
  \includegraphics[width = 8cm]{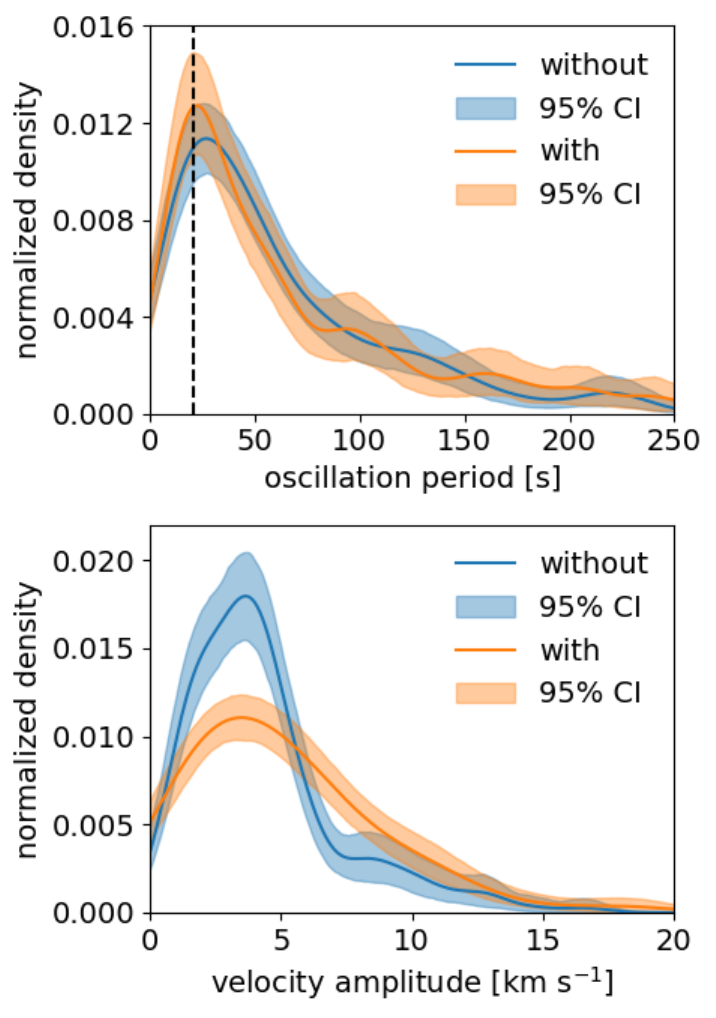}
  \caption{Same as Figure \ref{fig:hist_vel_prd}, but derived from horizontal velocities $v_{x}$ within the detected spicules.}
  \label{fig:hist_vel_prd_perp}
\end{figure}

Figure \ref{fig:hist_vel_prd_perp} shows the density distributions of velocity amplitudes and oscillation period of spicules similar to those shown in Figure \ref{fig:hist_vel_prd}, but obtained from horizontal velocities $v_{x}$ within the detected spicules. Table \ref{tab:oscil_tab_perp} presents the spicule oscillation parameters ($N_{\rm os}, \bar{P}_v$, and $\bar{A}_v$) derived from horizontal velocities. 


\begin{table*}[!h]
\centering
\caption{Same as Table \ref{tab:oscil_tab}, but derived from horizontal velocities $v_{x}$ within the detected spicules.}
\begin{tabular}{ccccccc}
\hline \hline 
\multicolumn{1}{c}{} & \multicolumn{3}{c}{without (mean $\pm$ standard error)} & \multicolumn{3}{c}{with (mean $\pm$ standard error)} \\
\cmidrule(rl){2-4} \cmidrule(rl){5-7} 
{height} & {$N_{\rm os, without}$} & {$\bar{P}_{v, \rm without}$} & {$\bar{A}_{v, \rm without}$} & {$N_{\rm os, with}$} & {$\bar{P}_{v, \rm with}$} & {$\bar{A}_{v, \rm with}$} \\
$(\rm km)$ & {} & {$(\rm s)$} & {$(\rm km \ s^{-1})$} & {} & {$(\rm s)$} & {$(\rm km \ s^{-1})$} \\
\hline
$4200$ & $213$ & $58 \pm 4$ & $4.5 \pm 0.2$ & $138$ & $67 \pm 6$ & $5.8 \pm 0.4$ \\
$4400$ & $209$ & $65 \pm 4$ & $4.7 \pm 0.2$ & $144$ & $61 \pm 5$ & $5.3 \pm 0.5$ \\
$4600$ & $186$ & $66 \pm 5$ & $4.3 \pm 0.2$ & $143$ & $71 \pm 5$ & $5.6 \pm 0.4$ \\
$4800$ & $171$ & $66 \pm 5$ & $4.3 \pm 0.2$ & $142$ & $71 \pm 5$ & $5.4 \pm 0.4$ \\
$5000$ & $135$ & $71 \pm 6$ & $4.5 \pm 0.3$ & $138$ & $73 \pm 5$ & $5.4 \pm 0.3$ \\
$5200$ & $125$ & $65 \pm 6$ & $3.9 \pm 0.3$ & $127$ & $71 \pm 6$ & $5.4 \pm 0.4$ \\
$5400$ & $100$ & $72 \pm 6$ & $4.3 \pm 0.3$ & $120$ & $68 \pm 6$ & $4.7 \pm 0.4$ \\
\hline
\end{tabular}
\label{tab:oscil_tab_perp}
\end{table*}

\clearpage






\end{document}